\documentstyle [12pt] {report}
\begin{document}
\hfuzz=50pt
\textheight 23.0cm
\topmargin -0.5in
%
%
%
%
\newcommand{\be}{\begin{equation}}
\newcommand{\ee}{\end{equation}}
\newcommand{\bea}{\begin{eqnarray}}
\newcommand{\eea}{\end{eqnarray}}
\begin{titlepage}
\begin{flushright}
IC/91/384 \\
November, 1991
\end{flushright}
\vspace{2cm}
\begin{center}
{ \large \bf Modified Black Holes in Two Dimensional Gravity }\\
 \vspace{2cm}

{\large\bf Noureddine Mohammedi}\footnote{nour@itsictp.bitnet}\\
\vspace{.5cm}
\large International Centre for Theoretical Physics \\
P. O. Box 586, 34100 Trieste, Italy.\\

\baselineskip 18pt

\vspace{.2in}

\vspace{1cm}
{\large\bf Abstract}
\end{center}
The $SL(2,R)/U(1)$ gauged WZWN model is modified by a topological term and
the accompanying change in the geometry of the two dimensional target space
is determined. The possibility of this additional term  arises from a
symmetry in the general formalism of gauging an isometry subgroup of
a non-linear sigma model with an antisymmetric tensor. It is shown, in
particular, that the space-time exhibits some general singularities for
which the recently found black hole is just a special case. From a conformal
field theory point of view and for special values of the unitary
representations of $SL(2,R)$, this topological term can be interpreted as a
small perturbation by a (1,1) conformal operator of the gauged WZWN action.\\

\end{titlepage}
\baselineskip 20pt

\setcounter{chapter}{1}
\setcounter{section}{1}
\setcounter{subsection}{1}
\subsection*{1.\ \,\,Introduction}
\setcounter{equation}{0}

Conformal field theories with central charge $c=1$, when coupled to
two-dimensional gravity provide interesting toy models for the study of string
theories. Perhaps the most appealing feature of these $c=1$ theories is
their ability to describe strings in non-critical dimensions in a
non-perturbative way [1-18]. This is due to the recent success of the matrix
model approach to non-critical strings. Another property of these theories is
their target space-time interpretation as critical string theories propagating
in non-trivial $1+1$  dimensional backgrounds [6-9].
In particular, the metric of
this two-dimensional target space exhibits a black hole singularity just
as the Schwarzschild black hole in four dimensions [19,20]. Of great
importance,
however, is the possibility of viewing these singularities as arising from
a modular invariant $SL(2,R)/U(1)$ coset WZWN model [19].
\par
In section two of this paper, we explore the change in the geometry of the
two dimensional target space due a modification by a topological term of
the gauged WZWN action. The possibility of this additional term stems from
the general formalism of gauging an arbitrary isometry subgoup of a non-linear
sigma model with an antisymmetric tensor. We find that, in general, the
scalar curvature is singular whenever the quantity
\be
{1\over 16(1-uv)^2}\left[4+2(uv-4)\partial_uX\partial_vX+4u\partial_uX
-4v\partial_vX+u^2(\partial_uX)^2+v^2(\partial_vX)^2\right]
\ee
has zeros or is undefined. Here $(u,v)$ are the target space coordinates and
$X(u,v)$ is a scalar function
introduced by the topological term.
\par
We also consider, in section three, the effects of the additional term on the
conformal field theory of the black hole as described by the $SL(2,R)/U(1)$
coset theory. We show that when the unitary representations $|l,m>$ of the
universal cover $\widetilde{SL}(2,R)$ and the level $k$ satisfy
\be
-{l(l+1)\over k-2}+{m^2\over 4k}=0\,\,\,,
\ee
the extra topological term is a conformal operator of dimension $(1,1)$.
Therefore it can be considered as a kind of perturbation of the ordinary
$SL(2,R)/U(1)$ coset model.

\setcounter{chapter}{2}
\setcounter{section}{1}
\setcounter{subsection}{1}
\subsection*{1.\ \,\,The Topological Term}
\setcounter{equation}{0}

The non-linear sigma model with a Wess-Zumino term  as given by
\be
S={1\over 2\pi}\int {\rm {d}}^2x\left(\sqrt {\gamma}\gamma^{\mu\nu}G_{ij}
+\epsilon^{\mu\nu}B_{ij}\right)\partial_\mu\phi^i\partial_\nu\phi^j
\ee
is invariant under the global $U(1)$ isometry transformations, ($\alpha$ is
constant),
\be
\delta\phi^i=\alpha K^i(\phi)\,\,\,,
\ee
provided that $K^i$ is a Killing vector of the metric $G_{ij}$,
$\left(\nabla_{(i}K_{j)}=0\right)$, and
\be
\partial_l H_{ijk}K^l +H_{ljk}\partial_iK^l+H_{ilk}\partial_jK^l
+H_{ijl}\partial_kK^l=0\,\,\,.
\ee
The torsion is defined by $H_{ijk}={3\over 2}\partial_{[i}B_{jk]}$. It
follows that the antisymmetric tensor $B_{ij}$ satisfies
\be
\partial_l B_{ij}K^l +B_{lj}\partial_iK^l+B_{il}\partial_jK^l=
\nabla_iL_j -\nabla_jL_i
\ee
for some vector $L_i$ [21].
\par
The above transformation can be made local, $\left(\alpha =\alpha (x)\right)$,
b
   y the introduction of a $U(1)$ gauge field $A_\mu$. Then the gauged action
takes the form [22,23]
\be
S_g={1\over 2\pi}\int d^2x\{\sqrt {\gamma}\gamma^{\mu\nu}G_{ij}
D_\mu\phi^iD_\nu\phi^j+\epsilon^{\mu\nu}B_{ij}
\partial_\mu\phi^i\partial_\nu\phi^j
-2\epsilon^{\mu\nu}C_iA_\mu\partial_\nu\phi^i\}\,\,\,,
\ee
where
\bea
D_\mu\phi^i&=& \partial_\mu\phi^i+A_\mu K^i\nonumber\\
C_i&=&B_{ij}K^j+L_i\,\,\,.
\eea
The gauge field $A_\mu$ transforms as $\delta A_\mu = -\partial_\mu \alpha$.
The
action (2.5) is then invariant under local gauge transformations if [22,23]
\bea
&\partial_jC_iK^j+C_j\partial_iK^j=0&\nonumber\\
&L_iK^i=0&\,\,\,.
\eea
\par
The gauge field apprears only quadratically and with no derivatives. Hence,
it can be eliminated via its equations of motion. The resulting theory is
again a non-linear sigma model with a new metric  $\widehat{G}_{ij}$ and an
antisymmetric tensor $\widehat{B}_{ij}$ given by
\bea
\widehat{G}_{ij}&=& G_{ij}-{1\over M}\left(G_{ik}G_{jl}K^kK^l + C_iC_j\right)
\nonumber\\
\widehat{B}_{ij}&=& B_{ij}+{1\over M}\left(G_{ik}C_jK^k -G_{jk}C_iK^k
\right)\,\
   ,\,,
\eea
where
\be
M=G_{ij}K^iK^j\,\,\,.
\ee
Notice that the new metric $\widehat{G}_{ij}$ would exhibit an explicit
singular
   ity
if $M$ has zeros. This is so if the old metric $G_{ij}$ is not
positive definite as it is in the case when the non-linear sigma model is
defined on a  non-compact group manifold. Using Eq.(2.8) we find
\be
\widehat{G}_{ij}K^j=0\,\,\,.
\ee
Therefore due to these null eigenvectors, the metric $\widehat{G}_{ij}$ cannot
be inverted and we cannot analyse the singularities of the gauged
non-linear sigma model. To overcome this difficulty,  a gauge fixing term
in the action (2.5) is necessary.
\par
The main remark in this note is that the defining equation for $L_i$, in (2.4),
is invariant under the shift [22]
\be
L_i\,\,\rightarrow \,\, L_i +\partial_i \lambda\,\,\,.
\ee
To guarantee that the gauged action (2.5) is  invariant under the above
shift, we should modify our action in the following manner
\be
S_g\,\,\rightarrow \,\,S_g + {1\over 2\pi}\int{\rm {d}}^2x\,\epsilon
^{\mu\nu}F_{\mu\nu}X(\phi)\,\,\,,
\ee
where $F_{\mu\nu}=\partial_\mu A_\nu-\partial_\nu A_\mu$ is the $U(1)$ field
strength. The additional term is gauge invariant if
\be
\partial_iXK^i=0\,\,\,.
\ee
The gauged action will be invariant under the shift (2.11) provided that
$X(\phi)$ undergoes the transformation
\be
X\,\,\rightarrow \,\, X +\partial_i \lambda\,\,\,.
\ee
Furthermore, the term in $F_{\mu\nu}$ is topological in nature.
\par
The possibility
of the additional term in Eq.(2.12) may also be seen from a different point of
view: In the normal coordinate expansion of the gauged action,
when the gauge field is taken as a fixed background, terms proportional
to $F_{\mu\nu}$ are generated [22]. Indeed, in order to expand the action $S_g$
around some classical field $\phi$ in a covariant fashion [24], we introduce
the quantum field $\xi^i(x)$ which is the tangent at $\phi^i$ to the
geodesic joining $\phi^i$ to $\phi^i+\pi^i$, where $\pi^i$ is a small
perturbation around $\phi^i$.  We find that the first term in the expansion
of the $\epsilon^{\mu\nu}$ term in $S_g$ is given by [22]
\be
{1\over 2\pi}\int d^2x\,\epsilon^{\mu\nu}\left(H_{ijk}D_\mu\phi^iD_\nu\phi^j
-C_kF_{\mu\nu}\right)\xi^k\,\,\,.
\ee
It is clear, therefore, that divergences
proportional to $F_{\mu\nu}$ will appear upon quantisation. Hence, for a
renormalisable theory we should add a term in $F_{\mu\nu}$ to the classical
action.
\par
To see the consequences of this additional term on the black hole physics,
let us apply our formalism
to the $SL(2,R)$ case.  The group manifold is parametrised by
\footnote{In this section we are using the conventions of ref.[19].}
\be
g=\left( \begin{array}{cc}
a&u\\
-v&{1\over a}(1-uv)
\end{array} \right)\,\,\,,
\ee
and the $U(1)$ isometry group is generated by
\footnote{Throughout this note we will restrict our analyses to gauging
the  non-compact $U(1)$ subgroup.}
\be
\delta a=2\alpha a\,\,\,, \delta u=\delta v=0\,\,\,.
\ee
The gauged WZWN action is given by an expression of the form (2.5),
(with the
factor ${1\over 2\pi}$ in the front replaced by $-{k\over 4\pi}$), and where
\bea
G_{ij}&=&\left(\begin{array}{ccc}
-{1\over a^2}(1-uv)&-{1\over 2}{v\over a}&-{1\over 2}{u\over a}\\
-{1\over 2}{v\over a}&0&{1\over 2}\\
-{1\over 2}{u\over a}&{1\over 2}&0
\end{array}\right )\nonumber\\
B_{ij}&=&\left(\begin{array}{ccc}
0&0&0\\
0&0&-\ln a\\
0&\ln a&0
\end{array}\right )\,\,\,.
\eea
Here $a=\phi^1\,,\,u=\phi^2\,,\,v=\phi^3$, and $L_i$ and $C_i$ are given
by
\bea
L_1&=&C_1=0\nonumber\\
L_2&=&C_2=2v\nonumber\\
L_3&=&C_3=-2u\,\,\,.
\eea
\par
Adding the term proportional to $F_{\mu\nu}$, as given in (2.12) with $X$
scaled by a factor $k$,  and integrating out the
gauge field leads to a non-linear sigma model with a metric $\widehat{G}_{ij}$
\bea
\widehat{G}_{uu}&=&{1\over
2(1-uv)}\left[-v\partial_uX-(\partial_uX)^2\right]\nonumber\\
\widehat{G}_{vv}&=&{1\over
2(1-uv)}\left[u\partial_vX-(\partial_vX)^2\right]\nonumber\\
\widehat{G}_{uv}&=&{1\over 2(1-uv)}\left[1+{1\over 2}u\partial_uX-
{1\over 2}v\partial_vX-\partial_uX\partial_vX\right]\,\,\,.
\eea
The coordinate $a$ has been eliminated via the gauge choice $a^2=1-uv$ [19].
Notice that by
setting  $X=0$ one gets the usual black hole metric [19,20].
\par
To illustrate the change in the geometry due to the $F_{\mu\nu}$ term, let us
compute the scalar curvature for a particular case for the function $X(u,v)$.
First of all we find, using (2.13) and (2.17), that $X$ is a function of the
coordinate
$u$ and $v$ only. Second of all, the only further restiction  that one
can have on $X$ is the requirement that it satisfies the equation of a scalar
field on the target space, namely
\bea
&\nabla^2X(u,v)=0&\nonumber\\
&=\left(-3u\partial_u-3v\partial_v+2(2-uv)\partial_u\partial_v
-u^2\partial_u^2-v^2\partial_v^2\right)X&\,\,\,,
\eea
where the covariant derivative $\nabla$ corresponds to the metric (2.18). The
last equation is also sufficient for the vanishing of the one-loop
counterterm proportional to $F_{\mu\nu}$ when the the gauge field is considered
as a fixed background. This counterterm is calculated from the expansion
in $\xi^i(x)$ given by
\be
{1\over 4\pi}\int d^2x\,\epsilon^{\mu\nu}F_{\mu\nu}\left(
\nabla_i\nabla_j-\nabla_iC_j\right)\xi^i\xi^j\,\,\,\,.
\ee
\par
The equation in (2.21) has two particular solutions
\bea
X_1(u,v)&=& xu^{-2}+yv^{-2}\nonumber\\
X_2(u,v)&=&z\ln ({u\over v})\,\,\,,
\eea
with $x$, $y$ and $z$ being some constants of integration. As an example,
the scalar curvature corresponding to the second solution is given by
\be
R_2={-4\left[u^2v^2(1+4z+6z^2+4z^3)+z^2(uv+1)(1+2z+2z^2)
\right]\over (1-uv)(uv+2zuv+2z^2)^2}\,\,.
\ee
In addition to the usual singularity at $uv=1$, there appear another
singularity at  $(uv+2zuv+2z^2)=0$. In general the extra singularities are
determined by
the curves in $u$ and $v$ for which the determinant of the metric
$\widehat{G}_{
   ij}$
is zero or undefined, as given in (1.1).

\setcounter{chapter}{3}
\setcounter{section}{1}
\setcounter{subsection}{1}
\subsection*{1.\ \,\,The Conformal Field Theory Of The Topological Term}
\setcounter{equation}{0}

In what follows we will investigate the effects of the additional term
involving $F_{\mu\nu}$ on the conformal field theory of the black hole
solution.
For this purpose the $SL(2,R)$ WZWN model is parametrised by
\footnote{In this section we will follow the notation of ref.[26].}
\be
g=e^{{i\over 2}\theta_L\sigma_2}e^{{1\over 2}r\sigma_1}
e^{{i\over 2}\theta_L\sigma_3}\,\,\,,
\ee
where $(r,\theta_L,\theta_R)$ are real Euler coordinates and $\sigma_i$ are the
Pauli matrices. The local gauge transformations correspond to
\be
\delta\theta_L=\delta\theta_R=\alpha\,\,,\,\,\delta r=0\,\,\,.
\ee
The gauged $SL(2,R)$ action takes the form
\bea
S=S_{wzwn}[r,\theta_L,\theta_R]&+&{k\over 2\pi}\int d^2z\left[A\left(
\bar {\partial}\theta_R+\cosh r\bar {\partial}\theta_L\right)\right.
\nonumber\\
&+&\left.\bar {A}\left(
\partial\theta_L+\cosh r\partial\theta_R\right)-\bar{A}A\left(\cosh r +1\right
)\right]\,,
\eea
where the  action $S_{wzwn}$ is given by
\be
S_{wzwn}={k\over 4\pi}\int d^2z\left(\bar{\partial} r\partial r
-\bar{\partial}\theta_L\partial\theta_L
-\bar{\partial}\theta_R\partial\theta_R-2 \cosh r
\bar{\partial}\theta_L\partial\theta_R\right)\,\,\,.
\ee
The gauge field $A=(A,{\bar{A}})$ is traded for two complex scalars
$\phi_L$ and $\phi_R$ ($\phi_L=\phi_R^{*}$) through [25]
\be
A=\partial\phi_L\,\,\,,\,\,\,{\bar{A}}={\bar {\partial}}\theta_R\,\,\,.
\ee
By redefining the fields as [26]
\be
\theta_L\rightarrow \theta_L+\phi_L,\,\,\,,\,\,\,
\theta_R\rightarrow \theta_R+\phi_R\,\,\,
\ee
one finds that the gauge fixed action has a dependence on $\phi_L$ and
$\phi_R$ through their difference $\phi=\phi_L-\phi_R$ only, and is given by
\be
S_{gf}=S_{wzwn}[r,\theta_L,\theta_R]+S[\phi]+S[b,c]\,\,\,,
\ee
where $S[\phi]$ describes a free scalar field
\be
S\left[\phi\right]=-{k\over 4\pi}\int d^2z\partial\phi\bar{\partial}\phi\,\,\,.
\ee
The Jacobian of the gauge fixing is given by a $(1,0)$ ghost system
represented by the action
\be
S\left[ b,c\right]=\int d^2z\left( b\bar{\partial}c+\bar{b}\partial
\bar{c}\right)\,\,\,.
\ee
\par
The energy momentum tensor corresponding to this action is given by
\be
T(z)={1\over k-2}\eta_{ab}J^aJ^b+{k\over 4}
\partial\phi\partial\phi+b\partial c\,\,\,,
\ee
and the $SL(2,R)$ currents are given by
\bea
J^3(z)&=&k\left(\partial \theta_L+\cosh r \partial\theta_R\right)\nonumber\\
J^\pm (z)&=&ke^{\pm i\theta_L}\left(\partial r\pm i\sinh r
\partial\theta_R\right)
\,\,\,.
\eea
The Virasoro algebra generated by $T(z)$ has a central charge
\be
c={3k\over k-2}-1\,\,\,.
\ee
The primary fields of this Euclidean coset theory have the form [26]
\be
T^l_{mn}(r,\theta_L,\theta_R)=P^l_{\omega_L\omega_R}(\cosh r)
e^{i\omega_L\theta_L+i\omega_R\theta_R}\,\,\,,
\ee
where the quantum numbers $\omega_L$ and $\omega_R$ are the eigenvalues
of $J_0^3$ and $\bar{J}_0^3$, respectively and $l$ labels the
$SL(2,R)$ isospin. The functions $P^l_{\omega_L\omega_R}$ are the Jacobi
functions, and
$\omega_L$ and $\omega_R$ take their values on the $m\times n$ lattice
\be
\omega_L={1\over 2}(m+nk)\,\,\,\,,\,\,\,\,\omega_R=-{1\over 2}(m-nk)\,\,\,.
\ee
The integers $m$ and $n$ are interpreted, respectively, as the discrete
momentum and the winding number of the string in the $\theta={1\over 2}
(\theta_L-\theta_R)$ direction [26].
\par
The vertex operators $T^l_{mn}$ are eigenfunctions of the Virasoro operators
$L_0$ and ${\bar{L}}_0$, which are represented by the differential operators
\bea
L_0&=&-{\triangle_0\over k-2}-{1\over k}{\partial\over\partial\theta^2_L}
\nonumber\\
{\bar{L}}_0&=&-{\triangle_0\over k-2}-{1\over k}{\partial
\over\partial\theta^2_R}\,\,\,\,,
\eea
with $\triangle_0$ being the Casimir (or the Laplacian) on the group manifold
and is given by
\be
\triangle_0={\partial^2\over\partial r^2}+\cosh r{\partial\over\partial r}
+{1\over\sinh^2 r}\left({\partial^2\over\partial \theta_L^2}-2\cosh r
{\partial^2\over\partial \theta_L\partial \theta_R}+
{\partial^2\over\partial \theta_R^2}\right)\,\,\,.
\ee
The eigenvalues of $L_0$ and ${\bar{L}}_0$ are the conformal weights of the
primary fields and are expressed as
\bea
h^l_{mn}&=&-{l(l+1)\over k-2}+{(m+nk)^2\over 4k}\nonumber\\
\bar{h}^l_{mn}&=&-{l(l+1)\over k-2}+{(m-nk)^2\over 4k}\,\,\,.
\eea
\par
The term in $F_{\mu\nu}$ that we want to add to our gauged action is given
by
\be
S_{add}={k\over\pi}\int d^2z\, X(r,\theta_L,
\theta_R){\bar{\partial}}\partial\phi\,\,\,.
\ee
On the other hand equation (2.13) yields
\be
\left(\partial_{\theta_L}+\partial_{\theta_R}\right)X=0\,\,\,.
\ee
Therefore
\be
X(r,\theta_L, \theta_R)=X(r,\theta_L-\theta_R)\,\,\,.
\ee
This additional term may be regarded as a perturbation to the conformal
field theory of the coset model. Hence, we require that
the operator ${\bar{\partial}}\partial
\phi X$ has conformal dimension $(1,1)$ with respect to the energy momentum
tensor of the unperturbed theory in (2.10). Since
${\bar{\partial}}\partial\phi$
is already of dimenion $(1,1)$ we must have
\be
L_0X={\bar{L}}_0X=0\,\,\,.
\ee
Therefore $X$ is a conformal operator of  the form (3.13) and of  dimension
$(0,0)$, and since it
depends only on the difference $\theta_L-\theta_R$ we deduce that the winding
number $n$ must be zero. Hence $X(r,\theta)$ is given by all the
vertex operators
\be
X(r,\theta)=P^l_{{1\over 2}m,-{1\over 2}m}(\cosh r)e^{{i\over
2}m(\theta_L-\theta_R)}\,\,\,,
\ee
for which the representations and the level $k$ have to satisfy
\be
-{l(l+1)\over k-2}+{m^2\over 4k}=0\,\,\,.
\ee
\par
To summarise, we have modified the gauged WZWN model by a term involving
the field strengh $F_{\mu\nu}$ of the $U(1)$ gauge group. We found that the
addition of this term changes completely the singularity structure of the
two-dimensional target space. We have calculated explicitly the scalar
curvature and found the singularity for a particular case of the additional
term. The conformal field theory of the black hole in the persence
of the $F_{\mu\nu}$ term was also analysed. In particular, this term can be
treated as a perturbation by a $(1,1)$ conformal operator of the $SL(2,R)/U(1)$
coset model.
\vspace{0.5cm}

{\bf Acknowledgements} : I would like to thank K. S.  Narain,
E. Gava, H. Sarmadi, S. Panda and B. Rai  for  many useful discussions. The
financial support from IAEA and UNESCO is also hereby acknowledged.

\end{document}